# On the Feasibility of Extreme Heating Rates in SEM using MEMS Heater Platforms

C. Koenig,[1] P. Mayr,[2] J.R. Jinschek,[1,*] A. Bastos Fanta[1,*]

[1] National Centre for Nano Fabrication and Characterization (DTU Nanolab), Technical University of Denmark (DTU), kgs. Lyngby, Denmark

[2] Chair of Materials Engineering of Additive Manufacturing, Technical University of Munich, Munich, Germany

* Corresponding authors: jojin@dtu.dk & absf@dtu.dk

**Abstract**

Understanding microstructural evolution under extreme thermal conditions is essential for advancing metal additive manufacturing (AM). This work demonstrates the feasibility of employing micro-electro-mechanical system (MEMS) heating platforms for in-situ scanning electron microscopy (SEM) characterization of bulk-like samples during rapid thermal cycling. Using electron backscatter diffraction (EBSD), we tracked the ferrite-to-austenite phase transformation in a pure iron specimen (≈ 100 × 100 × 5 µm) and confirmed that the sample surface temperature closely follows the MEMS temperature setpoint within device accuracy. Under vacuum conditions, stable heating and cooling rates of up to 1000 °C/s were achieved with minimal power input and without compromising EBSD pattern quality. These findings establish MEMS-based heating as a robust approach for in-situ microstructural characterization of AM-relevant thermal processes in the SEM, enabling quantitative studies of thermally activated phenomena such as diffusion, phase transformations, and microstructural evolution under far-from-equilibrium conditions.

## 1. Introduction

*In-situ* electron microscopy enables detailed investigations of the microstructural evolution of metals and alloys under operational thermal and heat treatment conditions, with spatial resolution across length scales from millimeters down to sub-nanometers. To facilitate such experiments, heating stages with stable and precise temperature control for integration into scanning electron microscopes (SEM) and transmission electron microscopes (TEM) have been developed over several decades. Coupling in-situ heating with advanced analytical techniques has enabled direct observation of dynamic processes such as phase transformations [1], [2], recrystallization [3], [4], [5], and thermal degradation [6], providing critical insights into the mechanisms that govern material behavior and performance.

Conventional SEM heating stages rely mainly on furnace-based systems, which are unsuitable for rapid thermal cycles and typically limited to heating and cooling rates of only a few °C per second [7], [8]. Operation at elevated temperatures also introduces challenges such as maintaining vacuum integrity, minimizing thermal gradients within the specimen, and managing thermal inertia [7], [8], [9], [10], [11]. Although furnace stages can reach temperatures up to 1300 °C, analytical techniques such as electron backscatter diffraction (EBSD) and energy-dispersive spectroscopy (EDS) are often restricted to below ~600 °C [12], [13] due to interference from infrared radiation (IR) and thermally induced image drift. These limitations motivate developing alternative in-situ heating strategies that provide thermal stability, minimal gradients, and substantially higher temperature ramping rates.

One such approach is based on Joule heating, in which an electrical current passes through a resistive material to generate heat directly. Early implementations used resistance-heated tantalum foils to heat metallic specimens for recrystallization studies [11], [14]. These studies assumed homogeneous temperature distributions and heating rates of ~100 °C/s, although such values were not experimentally verified. Building on the same physical principle, more recent developments have focused on miniaturized heating platforms, such as micro-electro-mechanical systems (MEMS)-based heaters, which are now widely used in *in-situ* TEM studies [15], [16]. MEMS devices are characterized by ultra-low thermal mass, localized heating, short thermal response times, minimal drift (<0.3 nm/min) [15], [16], [17]. These attributes enable exceptional thermal performance, including sustained operation above 1000 °C with power inputs of only tens of milliwatts and the capability for millisecond-scale thermal cycling [15], [17],

MEMS heaters were initially developed for electron-transparent samples (< 10 × 10 × 0.01 µm$^3$) *in-situ* studies in the TEM. However, recent studies have successfully adapted them for SEM environments [18]. Applications range from thin films [19] and micron-scale particles [20] to larger "slab"-like specimens (up to 50 × 40 × 20 µm), where recrystallization of copper and deformed titanium alloys has been studied [21], [22]. Moreover, these studies demonstrated reliable EBSD measurements at elevated temperatures, underscoring the feasibility of capturing dynamic processes in bulk-like samples.

Accurate knowledge of the temperature at the specimen surface during in-situ microscopy experiments is essential, since microstructural evolution depends on both the absolute temperature and the applied thermal gradients. Surface temperature uncertainty is particularly critical when extending MEMS heating applications from electron-transparent specimens (thickness <100 nm) to bulk-like samples (thickness in µm range). It is therefore essential to establish whether the sample surface temperature during experiments accurately corresponds to the MEMS heater setpoint. Uncertainty in the sample surface temperature complicates the interpretation of microstructural processes and limits comparability with complementary experimental techniques or theoretical models.

Establishing the correlation between heater surface and sample surface temperature is essential for applying MEMS-based heating to replicate complex and highly dynamic thermal environments, such as those involving rapid heating and cooling under far-from-equilibrium in metal additive manufacturing (AM) processes. AM processes involve extreme heating and cooling rates ($10^2$ - $10^6$ °C/s) [23], [24] suppressing equilibrium phase transformations and promoting metastable microstructures [25], [26]. While repeated layer-by-layer cycling drives solid-state transformations and produces hierarchical microstructures with chemical and physical heterogeneity across multiple length scales [25], [27], [28]. Therefore, replicating such far-from-equilibrium thermal conditions while obtaining quantitative in-situ data on thermal processes influencing the microstructure formation is critical for understanding microstructural evolution and tailoring AM processes.

The present work addresses the need to establish precise control over the experimental setup for MEMS-based heating systems for *in-situ* characterization of bulk-like samples. In contrast to the thin films and micron-scale particles previously studied, we focus on specimens with set dimensions of either 50 x 50x 5 µm or 100 × 100 × 5 µm, representing bulk-like material behavior. Our experiments, supported by COMSOL simulations, provide a calibration baseline for surface temperature accuracy, uniformity, and dynamic response. On this basis, we established a robust framework for applying MEMS-based heating in SEM studies of microstructural evolution, with direct relevance for AM and other processes involving extreme thermal cycling.

## 2. Experimental details

### 2.1. Finite element modelling

Finite element simulations (FEM) were performed in COMSOL Multiphysics® [29] to define sample size and estimate surface temperature, in-plane temperature uniformity, through-thickness gradients, and ramping rates. The simulations were based on the validated model of the DENSsolutions Wildfire® chip developed by Yang et al. [30], which couples electric currents, solid mechanics, and heat transfer in solids. Joule heating was simulated by applying an input voltage through the electric currents module, while heat dissipation was modeled via conduction and radiative losses in the in-plane (XY) direction. Thermal expansion was included to account for stresses caused by membrane bulging [17], and temperature coupling enabled the evaluation of the spatial temperature distribution. The device geometry and layer composition were reconstructed using light optical microscopy and SEM-based EDS.

Mesh discretization employed a free triangular mesh with element sizes ranging from 100 µm to 0.01 µm, a maximum growth rate of 1.3, a curvature factor of 0.6, and a narrow region resolution of 0.5. Mesh quality was evaluated in the original work by Yang [30] using skewness metrics and by comparing temperature distributions across different meshing conditions. Material properties were assigned from the COMSOL database [29]: Mo (138 W/m·K), SiN (20 W/m·K), and SiO (1.4 W/m·K). Simulated temperature distributions were validated against experimental measurements reported by van Omme et al. [17].

### 2.2. MEMS-heater stage for SEM

All experiments were performed using Wildfire® MEMS-based heating devices (DENSsolutions, Delft, The Netherlands), equipped with a silicon nitride (SiN) membrane and controlled via the DENS Digiheater software. For *in-situ* operation within the SEM, the chips were mounted on a custom-built holder with a flange feed-through, as illustrated in **Figure 1** [19]. This holder is compatible with various SEM stages, allowing experiments with different detectors such as EBSD and on-axis transmission Kikuchi diffraction (TKD) [19]. The stage used in this study allowed the bulk-like specimen and heating chip to be positioned either horizontally or at a pre-tilted angle of 45°, enabling EBSD analysis at elevated temperatures with a stage tilt of 25°.

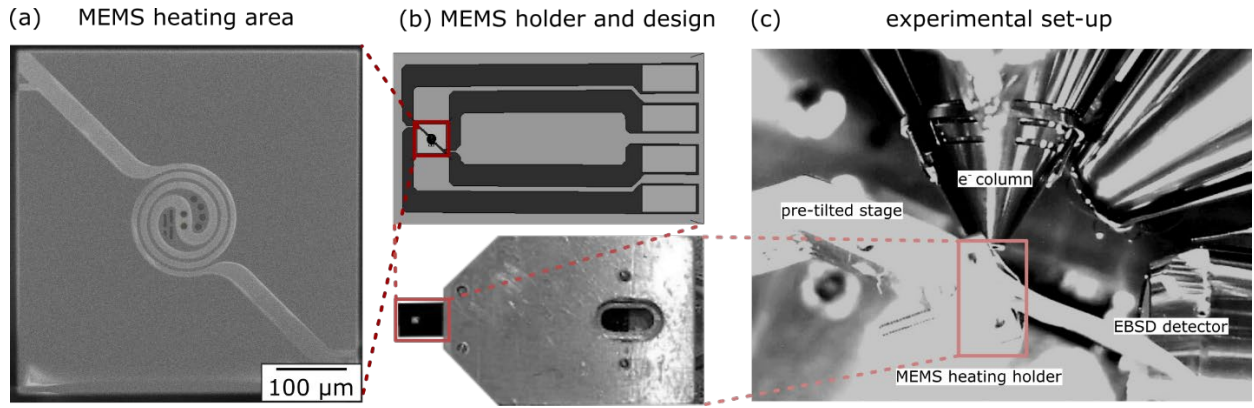

*Figure 1:* Overview of the MEMS-based heating setup for in-situ SEM experiments. (a) secondary electron image of the spiral-shaped molybdenum heating element. (b) schematic of the MEMS device [30] and optical image of the mounted chip on the customized stage. (c) SEM chamber view with MEMS holder on a pre-tilted stage and inserted EBSD detector

### 2.3. Sample preparation and lift out

The material investigated was high-purity Armco iron (99.85% Fe). Pure Fe was selected as a calibration standard due to its well-characterized ferrite (bcc) to austenite (fcc) phase transformation at approximately 910 °C, allowing indirect measure the surface temperature by following the phase transformation. Samples were mechanically polished using a Tegramin system (Struers A/S, Ballerup, Denmark) with a 0.25 µm diamond suspension, followed by final chemo-mechanical polishing with colloidal silica. The average grain size was on the order of 45 µm, suitable for EBSD investigations.

As the active heating area of the Wildfire® MEMS chips is limited to a diameter of 200 µm [17], EBSD was first performed on the bulk surface to identify regions of interest. From these areas, planar lift-out specimens of 100 × 100 µm or 50 × 50 µm were extracted using a Helios 5 Hydra UX plasma focused ion beam (PFIB) DualBeam® (Thermo Fisher Scientific Inc., Waltham, USA) operated with Xenon ions. This approach enabled direct comparison of data obtained from bulk material and PFIB-prepared regions, ensuring that high-energy ion milling did not introduce significant microstructural changes (see supplementary material S1).

The lift-out process followed the protocol described by Koenig et al. [31]. Trenches were milled around the selected region, and the block was extracted using an EasyLift needle with tungsten deposition. The backside was then polished flat to enable parallel alignment of the sample surface with the MEMS-heater. Finally, the specimen was centrally positioned on the MEMS device and secured using tungsten deposition applied to one side only. This limited, directional attachment minimized stress and prevented heater failure due to bulging at high temperatures [17]. Tungsten was deliberately chosen for this attachment due to its high melting point (~3400 °C [32]) and low diffusion rate in iron [33], which minimized potential surface contamination during heating.

### 2.4. Data acquisition and analysis

*In-situ* EBSD experiments were performed in a Helios 600 DualBeam® (Thermo Fisher Scientific Inc., Waltham, USA) equipped with an EDAX Hikari detector. The microscope was operated at 20 kV acceleration voltage, 22 nA beam current, and a 10 mm working distance. EBSD patterns were binned to a resolution of 96 × 96 pixels and indexed using a classic Hough-based algorithm with a low-resolution setting and a 9 × 9 convolution mask. A step size of 1° theta was used, and between 3 and 7 Kikuchi bands were detected per pattern. Datasets were processed using EDAX OIM Analysis 9™ software.

Two types of studies were carried out. The first focused on verifying the accuracy and uniformity of the sample surface temperature relative to the MEMS device $T_{set}$. For this purpose, EBSD maps were subjected to a comprehensive cleaning procedure, including a minimum confidence index (CI) threshold of 0.1, neighbor CI correlation, and single interaction grain dilation, with a grain tolerance angle of 5° and a minimum grain size of 5 pixels. The second study investigated heating and cooling rates using the time stamps of sequentially saved EBSD patterns. This enabled direct correlation of Digiheater-recorded temperature data with EBSD-derived quantities such as phase and pattern quality. In this case, only CI filtering with a 0.1 threshold was applied, without further cleaning.

## 3. Results and discussion

### 3.1. Influence of experimental conditions on temperature accuracy

The sample environment strongly influences the outcome of heating experiments, as it determines the relative contributions of heat conduction, convection, and radiation. These heat transfer mechanisms differ markedly between experiments performed in vacuum and in air, affecting not only the heating rate and temperature distribution but also the surface condition of the material through processes such as oxidation. Because these factors vary with sample environment, they also determine which approaches are suitable for calibrating the actual surface temperature of the sample. To evaluate how these environmental differences affect the performance of MEMS-based heaters during rapid thermal cycling, we performed comparative experiments under vacuum and atmospheric conditions using pristine Dens Solutions Wildfire chips with a silicon nitride membrane. Two target heating rates were selected, 100 °C/s and 1000 °C/s, the latter corresponding to the maximum achievable rate of the MEMS heater system. The actual heating rate was determined by fitting a linear slope to the measured temperature ($T_{meas}$) between 200 and 800 °C. The evolution of the set temperature ($T_{set}$), $T_{meas}$, and device power ($P_{heat}$) for both conditions is shown in **Figure 2**.

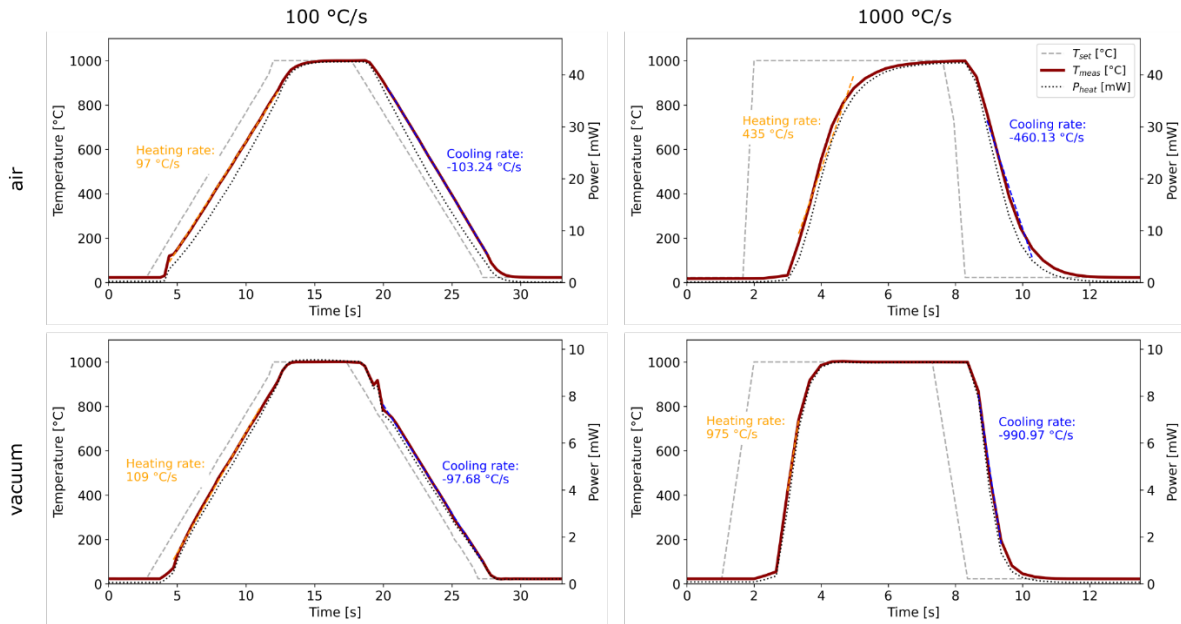

***Figure 2***: *Effect of environmental conditions on heating and cooling rates of MEMS heating device*

A consistent temporal delay of approximately 2 s between $T_{set}$ (dotted gray line) and $T_{meas}$ (solid red line) is observed across all temperature profiles. This delay remains uniform throughout the duration of each cycle, ensuring that the holding time at elevated temperatures corresponds to the specified duration. The proportional–integral–derivative (PID) control implemented in the software causes a leveling of the heating curve as the temperature approaches $T_{set}$. While this PID regulation prevents temperature overshoot and protects the MEMS device from potential damage, it also slightly reduces the nominal heating rate close to $T_{set}$.

At a heating rate of 100 °C/s, $T_{meas}$ closely follows $T_{set}$ in both air and vacuum, and the heating and cooling rates match the intended values. This consistency indicates that, at these moderate rates, the surrounding environment has only a minor effect on the thermal response of the device. In air, however, a noticeably higher power input ($P_{heat}$), up to 40 mW, was required to reach 1000 °C, compared to only 15 mW under vacuum conditions. When the target rate is increased to 1000 °C/s, a similar response is observed under vacuum, where the system successfully reaches the target temperature with power consumption comparable to that at the lower rate. In contrast, operation in air reveals clear limitations: the maximum achievable heating rate is restricted to approximately 435 °C/s, and the cooling rate to around 460 °C/s, both less than half of the intended values. The reduced rate also leads to a shorter duration at elevated temperature, which may alter the actual thermal exposure during experiments. These differences in heating behavior between air and vacuum are mainly caused by increased convective heat losses and the higher effective thermal mass of the surrounding atmosphere.

Consequently, when aiming for high heating rates, the MEMS device must be operated under vacuum conditions to minimize convective heat losses. While this enables stable and reproducible high-rate heating, it also constrains the experimental approach for calibrating the bulk-like samples surface temperature and the actual heating and cooling rates, as many conventional calibration techniques, such as thermo-cameras, cannot be applied under vacuum conditions [34].

Another key factor to ensure reliable experimental conditions during *in-situ* SEM heating is the interaction between the electron beam and the sample. Localized beam-induced heating can alter the sample or introduce uncertainties in the measured temperature, particularly when temperature is not actively regulated [34], [35]. Controlling these effects is therefore essential for precise thermal management. In our previous work [31], we investigated the influence of SEM acquisition parameters on beam-induced sample heating using the same MEMS setup employed in this study. A key finding was that the MEMS device's active four-point feedback loop dynamically adjusts the input power in response to beam-induced heating, stabilizing the temperature within approximately 3 s. This feedback mechanism allows the system to maintain stable thermal conditions even under high beam current conditions, such as those required for *in-situ* EBSD measurements. Consequently, although beam-induced heating is recognized as a contributing factor, the effectiveness of the feedback control ensures that its influence on the measured thermal response is negligible and therefore excluded from further discussion.

To further evaluate the stability of the experimental conditions, we examined the effect of sample temperature on the quality of the backscatter Kikuchi patterns. **Figure 3** shows a series of patterns acquired at increasing temperatures, following static background subtraction (background collected at room temperature). The first three patterns correspond to the ferritic (*bcc*) phase of iron, while the pattern obtained at 1000 °C reveals the expected high-temperature austenitic (*fcc*) structure. The image quality (IQ) values, indicating the sharpness and clarity of the patterns [36], remain nearly constant with increasing temperature, demonstrating that infrared (IR) radiation does not significantly affect detector performance. Similar results have been reported for in-situ TKD diffraction using MEMS heaters [19]. In contrast, previous *in-situ* SEM heating studies employing conventional furnace stages showed substantial IR interference with the detector, leading to reduced contrast in the diffraction pattern above 500–600 °C and requiring additional IR shielding [37]. The absence of such degradation in our experiments confirms that the small thermal mass of both the MEMS heater and the sample minimize IR emission, eliminating the need for specialized IR protection of the detector screen.

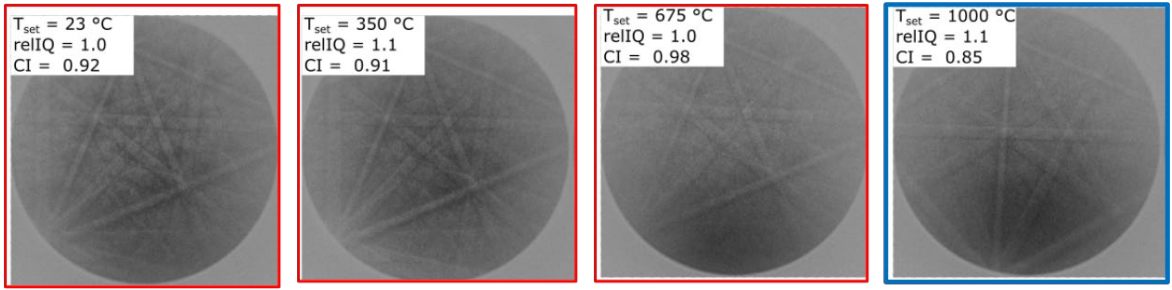

*Figure 3* Kikuchi patterns acquired at different $T_{set}$ values after static background subtraction using a reference collected at room temperature. Colors indicate the indexed phases: ferrite (bcc, red) and austenite (fcc, blue).

### 3.2. Assessing surface temperature uniformity in bulk-like samples

To achieve accurate temperature control in bulk-like samples surface, it is essential to define the region of the MEMS heater that provides uniform heating. For this purpose, in-plane FEM simulations of the temperature distribution and thermal response were performed over a range of input voltages [38]. **Figure 4a** shows the simulated temperature field across the heater area, and **Figure 4b** presents the corresponding temperature profile along the horizontal axis indicated in **Figure 4a**.

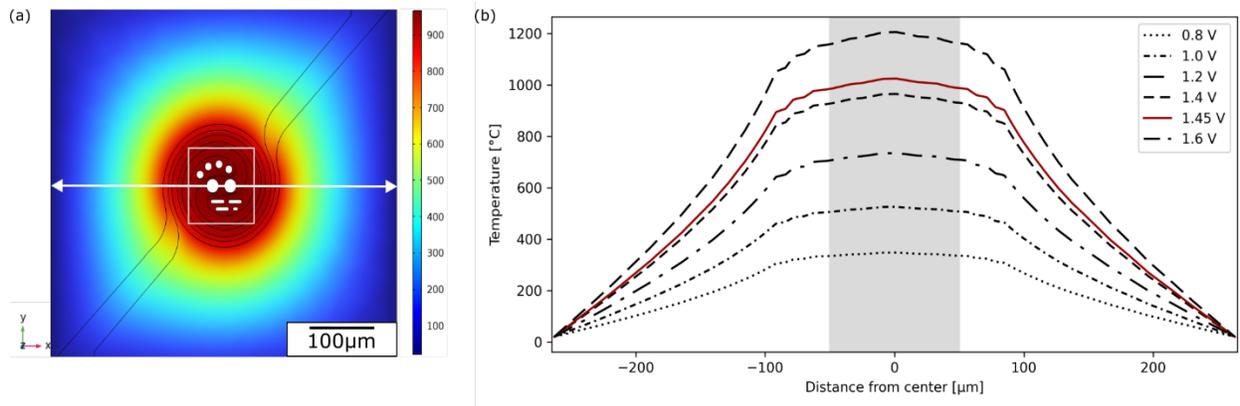

*Figure 4*: (a) Simulated temperature distribution in the XY-plane of the MEMS heater. White arrow indicates the horizontal line across the center used for temperature profile extraction, and the gray square highlights area of uniform temperature, as determined from (b). (b) Temperature profiles along the indicated line for various input voltages; gray area corresponds to the region of uniform temperature, adapted from [38].

The simulations show that the highest temperatures occur at the center of the MEMS heater, within the spiral core, while steep gradients develop toward the membrane edges. Increasing the input voltage raises the overall temperature and slightly changes the slope of these gradients. A central region of approximately 100 μm exhibits nearly uniform temperature, largely independent of input voltage. This region is shaded gray in **Figure 4b** and marked with a square in **Figure 4a**. Due to the symmetrical spiral design, the uniform temperature field extends along both x- and y-directions, consistent with the planar distribution shown in **Figure 4a**.

To evaluate how the presence of a specimen affects the heating behavior, a bulk-like iron sample (thermal conductivity 80.2 W/m·K, lateral dimensions 100 × 100 µm², thickness 5 µm) was included in the simulations, positioned above the coil center. **Figure 5** shows the simulated temperature distribution, including a top view (a), a cross-sectional side view (b), and temperature profiles from the sample center and corner (c).

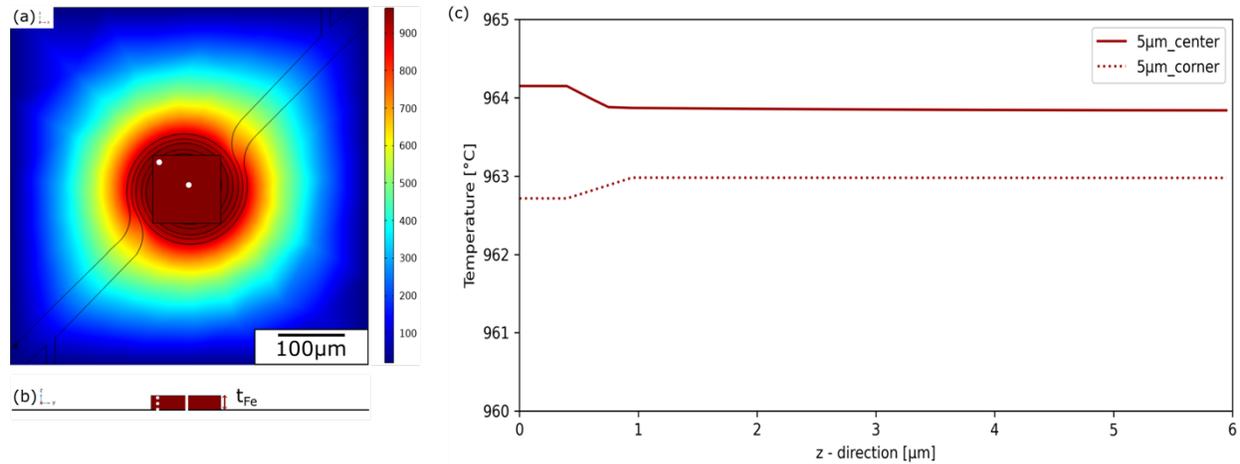

*Figure 5 Simulated temperature distribution shown as top (a) and cross-sectional views (b) for a 5 µm-thick iron sample placed on the MEMS heater. White markers in (a) and (b) indicate the positions used for temperature profile extraction, with center and corner profiles represented by solid and dotted lines, respectively. (c) Temperature profile along the thickness of the iron sample extracted from both center (solid lines) and corner (dotted lines) positions for an input voltage of 1.45 V, following [38].*

In the MEMS-heater-only configuration (**Figure 4**), an input voltage of 1.45 V produces a predicted heater temperature of 1000 °C. When the iron sample is included, the same input lowers the surface temperature to 963 °C at x = 0 µm, corresponding to the temperature at the interface heater-sample. The simulations further indicate that the vertical temperature drop between the heater bottom (x = 0 µm), and the sample surface (x = 6 µm) remains below 1 °C and occurs almost entirely within the first micrometer at the heater–sample interface (figure 5c). Above this region, the temperature is essentially uniform through the sample thickness. Only small lateral variations are predicted by the simulations: within the first micrometer, the center-corner difference reaches about 1.5 °C, but above 1 µm it decreases to less than 1 °C. These differences reflect local chip characteristics, since the center is supported by the molybdenum heating spiral, while the corners rest on the silicon nitride membrane without an active heater element. As shown experimentally by Vijayan et al. [39], the lower thermal conductivity of the membrane can locally reduce temperatures compared to the heating spiral setpoint. Nevertheless, the predicted temperature deviations remain well within the ±5 % uncertainty of the MEMS heater itself [17] and are therefore unlikely to affect measurements.

Overall, the simulations indicate that the high thermal conductivity of the bulk-like iron sample ensures efficient heat distribution and that the sample surface temperature closely follows the device temperature ($T_{set}$). At the same time, the presence of the bulk-like sample adds thermal load, requiring slightly higher input power ($P_{heat}$) to maintain the same target temperature as in the unloaded configuration. To verify this effect experimentally, $P_{heat}$ versus $T_{set}$ was measured for both a pristine chip and a chip with an attached iron sample (100 × 100 × 5 µm), as shown in **Figure 6**.

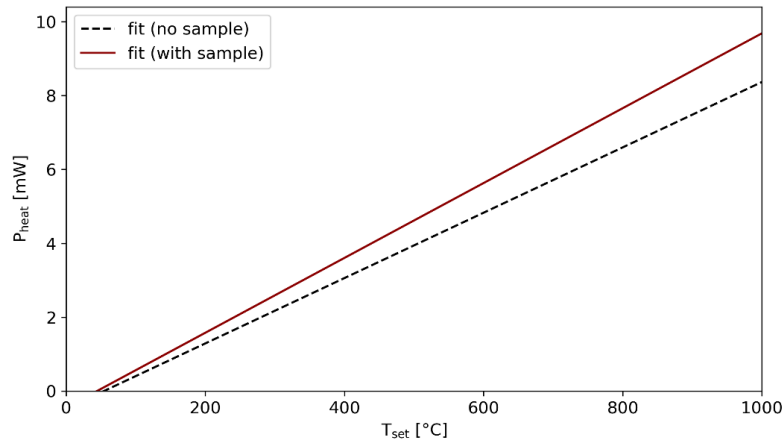

*Figure 6 Power required to reach set temperature $T_{set}$ of a MEMS chip under two conditions—pristine chip without a sample (dashed), and chip with a Fe sample (100 × 100 × 5 µm) (red).*

**Figure 6** shows that the system consistently requires a higher $P_{heat}$ when a sample is attached compared to the unloaded configuration. The two datasets were obtained from different MEMS chips; however, both originated from the same production batch and exhibited identical initial resistances of 162 Ω at room temperature, ensuring comparable baseline conditions. The observed increase in power demonstrates that the active feedback control of the MEMS heater detects the additional thermal load imposed by the sample and compensates by adjusting the input accordingly. As a result, the surface temperature achieved in both configurations can be considered equivalent within the accuracy limits of the MEMS heater.

While the simulations and power measurements indicate that the MEMS heater can maintain uniform and well-controlled temperatures even when loaded with a bulk-like sample, an experimental verification of the actual sample surface temperature was still required. For this purpose, the ferrite to austenite phase transformation in iron was used as an internal reference for temperature calibration because it provides a well-defined and reproducible benchmark for calibrating the sample surface temperature [40], [41]. The transformation temperature was first confirmed by differential scanning calorimetry (DSC) using a Netzsch STA 449 F1 Jupiter (NETZSCH-Gerätebau GmbH, Selb, Germany) at a controlled heating rate of 10 °C/min under an argon atmosphere (see supplementary material). For the *in-situ* experiments, a 5 µm-thick pure iron lift-out was attached to a Wildfire chip and EBSD measurements were performed to monitor the phase transition as a function of the MEMS $T_{set}$. After each completed map, $T_{set}$ was manually increased in 10 °C increments, with the Digiheater software applying the maximum available heating rate between steps.

Temperature uniformity was evaluated for two sample geometries, 50 × 50 µm × 5 µm and 100 × 100 µm × 5 µm, with acquisition times of about 11 s and 41 s per map, respectively. The corresponding phase maps (**Figure 7**) illustrate the transformation evolution with increasing $T_{set}$.

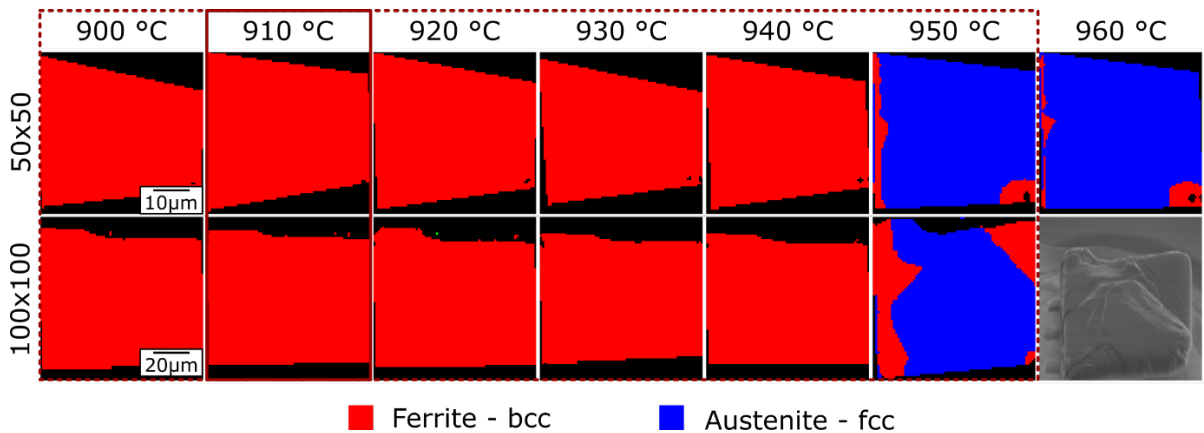

*Figure 7* EBSD phase maps acquired at varying $T_{set}$ for two sample sizes: 50 × 50 × 5 µm (top row) and 100 × 100 × 5 µm (bottom row). Colors indicate the corresponding crystallographic phases identified at each $T_{set}$.

As determined by DSC, the bcc-to-fcc transformation in iron occurs at 910 °C. Considering the specified ±5 % accuracy of the MEMS system [17], the transformation is therefore expected within a ±45 °C window around this temperature, as indicated by the dotted outline in **Figure 7**. In both samples, the transformation on the surface was observed at a $T_{set}$ of 950 °C, which lies within this temperature range of ±45 °C. Below this temperature, the material remains fully ferritic, and no early transformation is detected, indicating that the bulk-like sample surface follows a uniform thermal field. The sharp and well-defined transition shows that the temperature is homogeneous across the mapped region and that the surface temperature closely follows the MEMS heaters $T_{set}$ within its uncertainty. The simultaneous appearance of the *fcc* phase across both samples further suggests that lateral temperature gradients are negligible. The fact that both geometries, 50 × 50 µm × 5 µm and 100 × 100 µm × 5 µm, exhibit identical transformation behavior also confirms that heat transfer within the sample and across the interface is uniform and reproducible.

Localized regions that remained *bcc* above 950 °C are primarily found near the tungsten attachment sites. Tungsten was applied during the lift-out procedure, first to attach the Easylift needle (visible at the bottom right corner of the bulk-like sample) and later to secure the sample to the heater (as seen in the patches on the left side). The persistence of the *bcc* phase in these areas therefore likely reflects residual tungsten deposits, which retain their own *bcc* structure rather than indicating a true temperature variation across the sample surface. In addition, the upper edge of the larger 100 × 100 µm × 5 µm sample did not transform to the *fcc* iron structure. Secondary electron images recorded at 950 °C (**Figure 7**, lower right corner) show that the bulk-like sample had warped during the phase transformation and partially detached from the MEMS heater. This detachment was likely caused by thermal expansion of the polycrystalline material during the *bcc*-to-*fcc* transformation, which is accompanied by a volume increase and local strain accumulation at grain boundaries. Because the 5 µm-thick bulk-like sample was fixed only at discrete tungsten patches, these stresses were not evenly constrained, allowing portions of the sample to bend or lift out of plane. As a result, some regions lost direct thermal contact with the MEMS heater surface, reducing the efficiency of heat transfer across the interface. Such partial detachment can create local temperature gradients within the sample, which may slightly shift the apparent transformation temperature and contribute to the spatially heterogeneous phase distributions observed in the EBSD maps.

### 3.3. *Evaluating dynamic thermal response of bulk-like samples*

The previous results demonstrate that the MEMS heater provides accurate surface temperatures across bulk-like samples. Beyond temperature stability, however, a key advantage of MEMS-based heating devices lies in their ability to reach such conditions within milliseconds, enabling controlled rapid thermal cycling that is not achievable with conventional SEM heating stages. To assess this dynamic heating capability, time-dependent FEM simulations were performed. A fully coupled thermal model was implemented with a time step of 0.01 s, while all other parameters remained consistent with the stationary simulations discussed above. The analysis focused on how quickly a uniform temperature field develops across the surface of the bulk-like sample. Temperature changes were monitored at both the center and corner positions of the sample. **Figure 8** shows the simulated temperature distributions at two selected time points in planar view (a), along with the corresponding temporal temperature profiles for a 5 µm-thick sample (b).

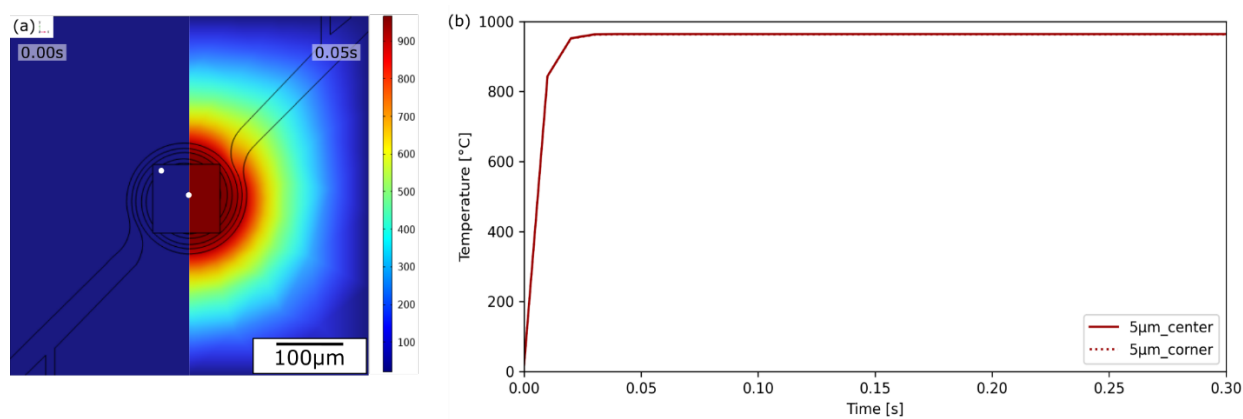

*Figure 8* Simulated temperature distribution shown of the MEMS heater and bulk-like sample at 0.00 s (a) and 0.05 s (b). White markers indicate the positions used for temperature profile extraction, with center and corner locations corresponding to solid and dotted lines, respectively. (c) Temperature profiles over time, extracted from both center (solid lines) and corner (dotted lines) positions

The simulations show that the surface temperature of the bulk-like iron sample increases rapidly within the first few tens of milliseconds and reaches a stable value after approximately 0.1 s. The temperature profiles at the center and corner positions overlap throughout the entire heating process, indicating that the surface reaches thermal equilibrium uniformly and without delay. Once equilibrium is established, the resulting temperature distribution matches the stationary simulations presented earlier, confirming that steady-state calculations accurately represent the thermal conditions during constant heating. The simulations predict that the sample surface closely follows the MEMS heater's $T_{set}$, with uniform heating conditions established almost instantaneously and remaining stable once the target temperature is reached.

To experimentally confirm these findings and to evaluate the surface temperature response of the bulk-like sample under rapid thermal cycling, the same approach based on the *bcc*-to-*fcc* phase transformation in iron was applied. Controlled heating experiments were performed on a bulk-like iron lift-out (50 × 50 × 5 µm) mounted on the Wildfire chip. The heating rates were varied from 100 °C/s to 1000 °C/s, with temperature profiles programmed in the Digiheater software. Each profile reached a peak $T_{set}$ of 1000 °C, followed by a 5 s hold and subsequent cooling to room temperature at the same rate. EBSD data were collected at 250 patterns per second (4 ms exposure time), providing high temporal resolution of the phase evolution. The $T_{set}$ and $T_{meas}$ values from the MEMS device were recorded every 0.3 s, allowing direct correlation with the phase maps using time as the common reference. Each EBSD measurement was performed on a 7 × 7 µm² subarea at the center of a single grain to minimize the influence of grain boundaries on the transformation. The results are presented in **Figure 9**.

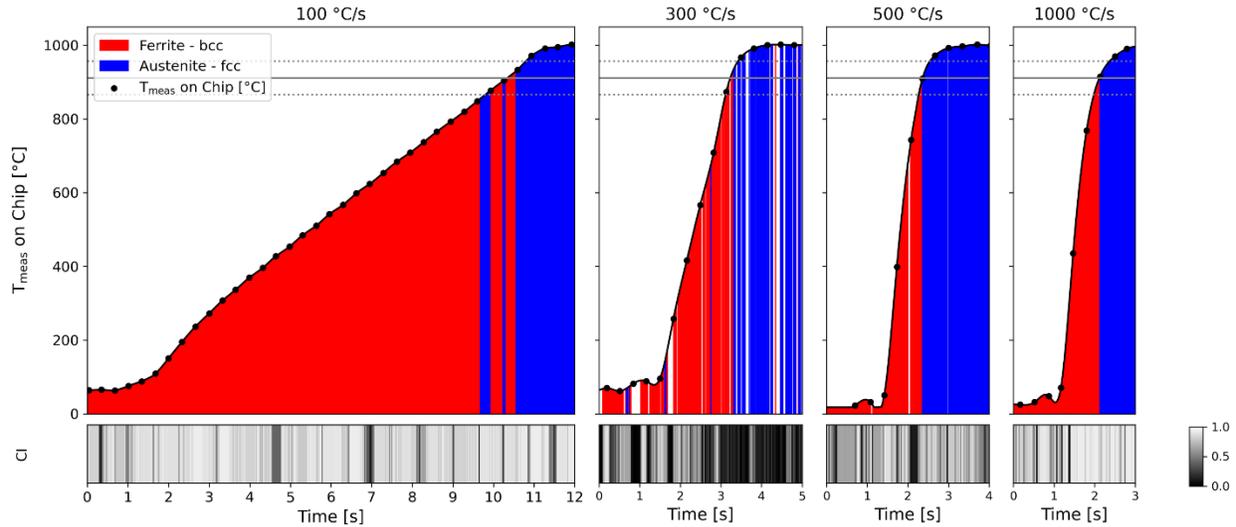

**Figure 9** $T_{meas}$ and phase identification by EBSD during heating at 100, 300, 500, and 1000 °C/s (left to right). Red and blue regions indicate ferrite (bcc) and austenite (fcc) phases, respectively. Black dots show $T_{meas}$; grayscale bars below represent CI over time. Plot widths are scaled uniformly, so one second corresponds to the same horizontal length in each subplot, enabling direct time-based comparison. Horizontal lines indicate the reference phase transformation temperature determined by DSC with dotted lines representing the 5% accuracy bounds.

At a heating rate of 100 °C/s, the phase transformation on the bulk-like sample surface was completed at a temperature of about 950 °C, which agrees well with the transformation temperature determined in the temperature uniformity study (see **Figure 7**). Between 850 °C and 950 °C, intermittent phase switching was observed, suggesting a temporary coexistence of the *bcc* and *fcc* phases. Such behavior has been attributed to a dynamic transformation front, where local strain accumulation or bursts of nucleation influence the transformation kinetics [40]. Langner et al. [41] similarly reported that even during rapid transformations, short-lived coexistence of both phases can occur along the advancing interface.

At a heating rate of 300 °C/s, a comparable transformation sequence was observed, with the completion of the transformation around 960 °C. The overall CI values were slightly lower, as indicated by the darker grayscale contrast, which may result from surface relief formation during the transformation or from a locally degraded surface finish of the lift-out. Both effects can cause shadowing and reduce pattern quality; however, all CI values remained above 0.1s, confirming reliable indexing of the EBSD data throughout the measurement.

At a heating rate of 500 °C/s, the transformation temperature decreased slightly to approximately 900 °C, still within the ±5 % temperature accuracy range of the MEMS heater. This shift suggests that the minor offset observed between $T_{set}$ and the actual transformation temperature likely originates from the intrinsic uncertainty of the system rather than from the thermal response of the bulk-like sample.

At the highest heating rate of 1000 °C/s, the transformation was observed at 910 °C, matching the temperature determined independently by DSC for bulk iron.

The close agreement across all heating rates confirms that the surface temperature of the bulk-like sample surface temperature accurately follows the programmed thermal profile of the MEMS heater, even under rapid heating conditions. With increasing heating rate, the number of discontinuous phase switches between the *bcc* and *fcc* structures decreased. This is consistent with the findings of Langner et al. [41], who reported that slower heating rates extend the overall transformation time, increasing the likelihood of capturing intermediate or mixed-phase states. In contrast, faster heating rates promote a more abrupt phase transition, reducing both the temporal window of phase coexistence and the number of diffraction patterns acquired within each temperature interval, since the total experiment duration was kept constant. Furthermore, the experiment at 100 °C/s was performed on a different lift-out than the higher-rate experiments, yet both exhibited equivalent transformation behavior, confirming the reproducibility of the heating response across independent bulk-like samples.

Following the analysis of the heating behavior, the cooling response of the system was investigated to determine whether the phase transformation and temperature tracking during cooling exhibit comparable characteristics. These measurements were conducted under the same experimental conditions as the heating experiments, using the same bulk-like iron lift-outs and equivalent thermal profiles defined by the Digiheater software. The results for the different controlled cooling rates are presented in **Figure 10**.

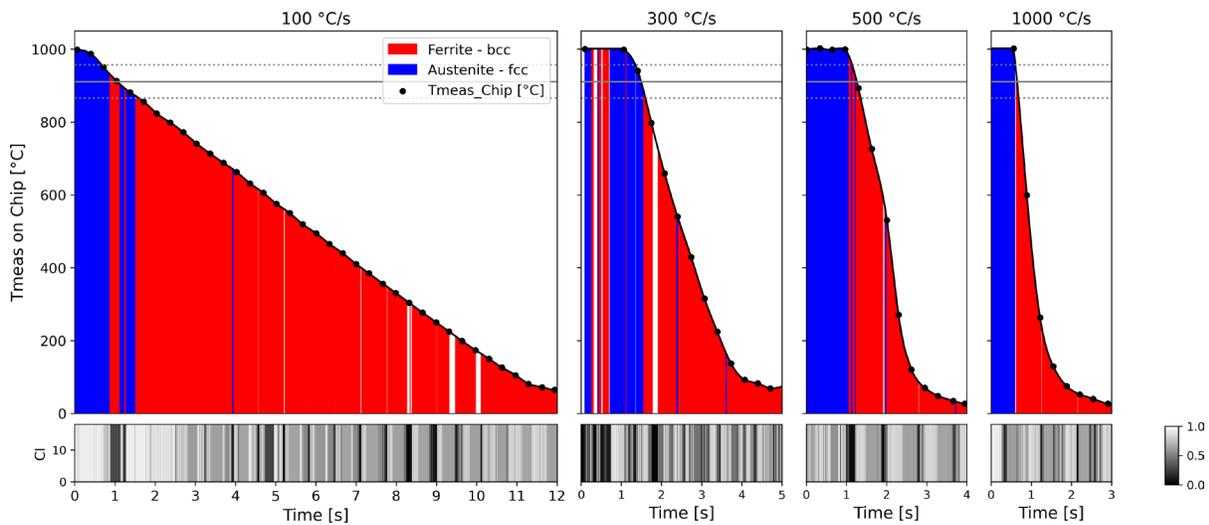

***Figure 10***: $T_{meas}$ *and phase identification by EBSD during cooling at 100, 300, 500, and 1000 °C/s (left to right). Red and blue regions indicate ferrite (bcc) and austenite (fcc) phases,*

*respectively. Black dots show $T_{meas}$; grayscale bars below represent CI over time. Plot widths are scaled uniformly, so one second corresponds to the same horizontal length in each subplot, enabling direct time-based comparison. Horizontal lines indicate the reference phase transformation temperature determined by DSC with dotted lines representing the 5% accuracy bounds.*

The phase transformation behavior as a function of time and $T_{meas}$ closely reflects the trends observed during heating. In all cases, the transformation occurred within the ±5 % temperature accuracy range of the MEMS heater, independent of the applied cooling rate. A slightly lower phase transformation temperature was observed during the reverse transformation from *fcc* to *bcc*, which is consistent with the expected thermal hysteresis of the allotropic transition in iron. For a cooling rate of 100 °C/s, the transformation completed at approximately 860 °C; at 300 °C/s, around 870 °C; at 500 °C/s, approximately 905 °C; and at 1000 °C/s, around 915 °C, respectively. The minor phase transformation temperature variations across rates likely reflect small differences in the local thermal gradients or in the nucleation dynamics of the reverse transformation rather than a systematic deviation in temperature control. Although the MEMS heater does not include active cooling, the small thermal mass of the chip and the SEM vacuum environment enable efficient heat dissipation, allowing cooling rates up to 1000 °C/s to be achieved reproducibly. The *fcc* to *bcc* phase transformation proceeded uniformly across the mapped region, with no evidence of delayed phase reformation or localized temperature lag. These results further demonstrate that the surface temperature of the bulk-like sample follows the programmed thermal response of the MEMS heater during both heating and cooling. The good agreement between experimental and simulated data confirms that the bulk-like sample surface accurately follows both the absolute $T_{set}$ and the applied thermal rate, validating the approach for studying rapid phase transformations under controlled conditions.

### 4. Summary and conclusions

This study demonstrates the feasibility of MEMS-based heating platforms for *in-situ* SEM investigations of bulk-like metallic samples subjected to rapid thermal cycling. Simulations and experiments consistently show that the surface temperature of samples up to 100 × 100 × 5 µm closely follows the programmed temperature of the MEMS heater, even during heating and cooling rates up to 1000 °C/s.

The temperature response of the MEMS system is affected by the experimental environment: in air, convective losses increase the required power input and limit the achievable heating rates, whereas under vacuum the device operates efficiently and reproducibly. The integrated four-point feedback loop dynamically regulates the power input in response to changes in thermal load, including those introduced by the sample and by electron beam irradiation [31], thereby maintaining stable and controlled temperature profiles throughout *in-situ* measurements. The small thermal mass of both heater and sample minimizes drift and radiative losses, allowing analytical techniques such as EBSD to be performed reliably across the full temperature range.

The accuracy of surface temperature measurements obtained by tracking the phase transformation of pure iron is inherently limited by the accuracy of the MEMS heating device itself and cannot be decoupled from it. Similar accuracy has previously been reported [17], confirming that for highly conductive, bulk-like samples, the surface temperature remains uniform and aligns well with both numerical predictions. This result indicates that the temperature measured by the MEMS device closely represents the actual sample surface temperature for highly conductive bulk-like samples. In contrast, materials with low thermal conductivity are expected to exhibit slower temperature response and increased local gradients. The agreement between stationery and time-dependent simulations as well as experimental observations further demonstrates that temperature equilibrium across the sample is established within fractions of a second, enabling precise control even under rapid thermal cycling.

This systematic study establishes a reliable framework for applying MEMS-based heating in the SEM to investigate microstructural evolution under rapid and well-defined thermal conditions. While the present work is limited to sample volumes of up to 100 × 100 × 5 µm, this already exceeds the size range typically accessible in TEM-based MEMS experiments. The results suggest that heating experiments using even larger, highly conductive samples may be feasible, provided that MEMS chip integrity and mechanical stability are maintained. The insights gained here provide a foundation for extending MEMS-based *in-situ* SEM studies to larger-scale and more complex materials, enabling high-rate investigations of thermally activated processes with improved control and quantitative accuracy.

**CRediT authorship contribution statement**

**C. Koenig:** Conceptualization, methodology, formal analysis, investigation, data curation, writing original draft, writing review & editing. **Peter Mayr:** Supervision, funding acquisition, writing review & editing **J.R. Jinschek:** Conceptualization, writing review & editing, supervision, funding acquisition **A. Bastos Fanta:** Conceptualization, methodology, writing review & editing, supervision

**Declaration of competing interests**
The authors declare that they have no known competing financial interest or personal relationships that could have appeared to influence the work reported in this paper.

**Data availability:**
The data that supports the findings of this study are available from the corresponding author, upon reasonable request.

**Acknowledgements**

The authors acknowledge the support of their colleagues at the National Centre for Nano Fabrication and Characterization in Denmark (DTU Nanolab) at the Technical University of Denmark (DTU) for their scientific contributions and the many fruitful discussions. Financial support from DTU, enabling a DTU Alliance project in collaboration with Technical University of Munich, is gratefully acknowledged.

**References**

[1] I. Lischewski, D. M. Kirch, A. Ziemons, and G. Gottstein, "In-situ EBSD study of the α-γ-α phase transformation in a microalloyed steel," *Ceramic Transactions*, vol. 201, pp. 95–102, 2008, doi: 10.1002/9780470444214.CH10.

[2] G. G. E. Seward *et al.*, "High-temperature electron backscatter diffraction and scanning electron microscopy imaging techniques: In-situ investigations of dynamic processes," *Scanning*, vol. 24, no. 5, pp. 232–240, 2002, doi: 10.1002/SCA.4950240503.

[3] G. J. Liao, R. Le Gall, and G. Saindrenan, "Experimental investigations into kinetics of recrystallisation of cold rolled nickel," *Materials Science and Technology*, vol. 14, no. 5, pp. 411–416, 1998, doi: 10.1179/MST.1998.14.5.411.

[4] S. Taylor, I. Masters, Z. Li, and H. R. Kotadia, "Direct Observation via In Situ Heated Stage EBSD Analysis of Recrystallization of Phosphorous Deoxidised Copper in Unstrained and Strained Conditions," *Metals and Materials International*, vol. 26, no. 7, pp. 1030–1035, Jul. 2020, doi: 10.1007/S12540-019-00493-Y.

[5] Y. Huang, J. He, X. Gu, G. Xu, Q. Qi, and J. Luo, "In situ EBSD study on the abnormal grain growth phenomenon in high-purity cobalt," *J Mater Sci*, vol. 59, no. 39, pp. 18698–18711, Oct. 2024, doi: 10.1007/S10853-024-10271-W.

[6] C. Y. Chen *et al.*, "In situ Scanning Electron Microscopy of Silicon Anode Reactions in Lithium-Ion Batteries during Charge/Discharge Processes," *Sci Rep*, vol. 6, Oct. 2016, doi: 10.1038/SREP36153.

[7] R. Heard, J. E. Huber, C. Siviour, G. Edwards, E. Williamson-Brown, and K. Dragnevski, "An investigation into experimental in situ scanning electron microscope (SEM) imaging at high temperature," *Review of Scientific Instruments*, vol. 91, no. 6, Jun. 2020, doi: 10.1063/1.5144981.

[8] J. Mendonça, H. P. Brau, D. Nogues, A. Candeias, and R. Podor, "Development of a microfurnace dedicated to in situ scanning electron microscope observation up to 1300 °C. I. Concept, fabrication, and validation," *Review of Scientific Instruments*, vol. 95, no. 5, May 2024, doi: 10.1063/5.0207466.

[9] G. G. E. Seward *et al.*, "High-temperature electron backscatter diffraction and scanning electron microscopy imaging techniques: In-situ investigations of dynamic processes," *Scanning*, vol. 24, no. 5, pp. 232–240, Sep. 2002, doi: 10.1002/SCA.4950240503.

[10] M. Bestmann, S. Piazolo, C. J. Spiers, and D. J. Prior, "Microstructural evolution during initial stages of static recovery and recrystallization: new insights from in-situ

heating experiments combined with electron backscatter diffraction analysis," *J Struct Geol*, vol. 27, no. 3, pp. 447–457, Mar. 2005, doi: 10.1016/J.JSG.2004.10.006.

[11] N. Bozzolo, S. Jacomet, and R. E. Logé, "Fast in-situ annealing stage coupled with EBSD: A suitable tool to observe quick recrystallization mechanisms," *Mater Charact*, vol. 70, pp. 28–32, Aug. 2012, doi: 10.1016/J.MATCHAR.2012.04.020.

[12] H. Mansour *et al.*, "High-temperature EDS and EBSD Analysis – Enabling In Situ Heating for Direct Observation of Phase Transformations in the SEM," *Microscopy and Microanalysis*, vol. 29, no. Supplement_1, pp. 2085–2086, Jul. 2023, doi: 10.1093/MICMIC/OZAD067.1079.

[13] M. Fekih, T. M. Ostormujof, L. Germain, M. Piette, D. Sornin, and N. Gey, "High temperature EBSD experiment versus crystallographic reconstruction to study phase transformation induced microstructures - Application examples for steels," *Mater Charact*, vol. 228, p. 115427, Oct. 2025, doi: 10.1016/J.MATCHAR.2025.115427.

[14] G. J. Liao, R. Le Gall, and G. Saindrenan, "Experimental investigations into kinetics of recrystallisation of cold rolled nickel," *Materials Science and Technology*, vol. 14, no. 5, pp. 411–416, 1998, doi: 10.1179/MST.1998.14.5.411.

[15] L. Mele *et al.*, "A MEMS-based heating holder for the direct imaging of simultaneous in-situ heating and biasing experiments in scanning/transmission electron microscopes," *Microsc Res Tech*, vol. 79, no. 4, pp. 239–250, Apr. 2016, doi: 10.1002/JEMT.22623.

[16] L. F. Allard, W. C. Bigelow, M. Jose-Yacaman, D. P. Nackashi, J. Damiano, and S. E. Mick, "A new MEMS-based system for ultra-high-resolution imaging at elevated temperatures," *Microsc Res Tech*, vol. 72, no. 3, pp. 208–215, Mar. 2009, doi: 10.1002/JEMT.20673.

[17] J. T. van Omme, M. Zakhozheva, R. G. Spruit, M. Sholkina, and H. H. Pérez Garza, "Advanced microheater for in situ transmission electron microscopy; enabling unexplored analytical studies and extreme spatial stability," *Ultramicroscopy*, vol. 192, pp. 14–20, Sep. 2018, doi: 10.1016/J.ULTRAMIC.2018.05.005.

[18] A. Bondarev, P. Sreenivasa Rao, S. Isavand, and N. O'Dowd, "Revealing the reverse austenite and martensite transformation pathways in P91 steel through in situ MEMS-heating EBSD investigations," *Mater Today Commun*, vol. 43, Feb. 2025, doi: 10.1016/j.mtcomm.2025.111753.

[19] A. B. Fanta *et al.*, "Elevated temperature transmission Kikuchi diffraction in the SEM," *Mater Charact*, vol. 139, pp. 452–462, May 2018, doi: 10.1016/J.MATCHAR.2018.03.026.

[20] N. Erdman, T. Laudate, and S. Mick, "Time-Resolved EDS Studies with Rapid Heating and FEG-SEM," *Microscopy and Microanalysis*, vol. 17, no. S2, pp. 514–515, Jul. 2011, doi: 10.1017/S1431927611003448.

[21] L. Novák, J. Stárek, T. Vystavěl, and L. Mele, "MEMS-based Heating Element forin-situDynamical Experiments on FIB/SEM Systems," *Microscopy and Microanalysis*, vol. 22, no. S3, pp. 184–185, Jul. 2016, doi: 10.1017/S143192761600177X.

[22] M. Wu and L. Novak, "Direct Real Time Microstructure Evolution Observation of Ti-6Al-4V Alloy by In-Situ EBSD during Cycling Heating and Cooling at Elevated Temperatures using heater in SEM/FIB Systems," *Microscopy and Microanalysis*, vol. 24, no. S1, pp. 818–819, Aug. 2018, doi: 10.1017/S1431927618004580.

[23] U. Scipioni Bertoli, G. Guss, S. Wu, M. J. Matthews, and J. M. Schoenung, "In-situ characterization of laser-powder interaction and cooling rates through high-speed imaging of powder bed fusion additive manufacturing," *Mater Des*, vol. 135, pp. 385–396, Dec. 2017, doi: 10.1016/J.MATDES.2017.09.044.

[24] W. Liu, G. Hu, Z. Yan, B. Liu, T. Wang, and Z. Lyu, "Thermal effect on microstructure and mechanical properties in directed energy deposition of AISI 316L," *International Journal of Advanced Manufacturing Technology*, vol. 134, no. 7–8, pp. 3337–3353, Oct. 2024, doi: 10.1007/S00170-024-14274-4.

[25] E. Toyserkani, D. Sarker, O. Obehi Ibhadode, F. Liravi, P. Russo, and K. Taherkhani, "Metal additive manufacturing," *Metal Additive Manufacturing*, pp. 1–594, Oct. 2021, doi: 10.1002/9781119210801.

[26] M. Ramoni, R. Shanmugam, N. Thangapandian, and M. Vishnuvarthanan, "Challenges in Additive Manufacturing for Metals and Alloys," *Springer Tracts in Additive Manufacturing*, vol. Part F3251, pp. 57–72, 2024, doi: 10.1007/978-3-030-89401-6_3/FIGURES/12.

[27] G. M. Karthik and H. S. Kim, "Heterogeneous Aspects of Additive Manufactured Metallic Parts: A Review," *Metals and Materials International*, vol. 27, no. 1, pp. 1–39, Jan. 2021, doi: 10.1007/S12540-020-00931-2/FIGURES/30.

[28] S. Gaudez *et al.*, "High-resolution reciprocal space mapping reveals dislocation structure evolution during 3D printing," *Addit Manuf*, vol. 71, p. 103602, Jun. 2023, doi: 10.1016/J.ADDMA.2023.103602.

[29] "COMSOL - Software for Multiphysics Simulation." Accessed: Sep. 29, 2025. [Online]. Available: https://www.comsol.com/

[30] Y.-C. Yang, "Establishment of controlled thermal conditions in in-situ TEM heating experiments: Calibration and Application," 2025. Accessed: Sep. 29, 2025. [Online]. Available: https://orbit.dtu.dk/en/publications/establishment-of-controlled-thermal-conditions-in-iin-situi-tem-h

[31] C. Koenig, A. B. da S. Fanta, and J. R. Jinschek, "Measurement of electron beam induced sample heating in SEM experiments," *Ultramicroscopy*, vol. 276, p. 114195, Oct. 2025, doi: 10.1016/J.ULTRAMIC.2025.114195.

[32] I. Langmuir, "The Melting-Point of Tungsten," *Physical Review*, vol. 6, no. 2, p. 138, Aug. 1915, doi: 10.1103/PhysRev.6.138.

[33] S. Takemoto, H. Nitta, Y. Iijima, and Y. Yamazaki, "Diffusion of tungsten in α-iron," *Philosophical Magazine*, vol. 87, no. 11, pp. 1619–1629, Apr. 2007, doi: 10.1080/14786430600732093.

[34] F. Gaulandris, S. B. Simonsen, J. B. Wagner, K. Mølhave, S. Muto, and L. T. Kuhn, "Methods for Calibration of Specimen Temperature During In Situ Transmission Electron Microscopy Experiments," *Microscopy and Microanalysis*, vol. 26, no. 1, pp. 3–17, Feb. 2020, doi: 10.1017/S1431927619015344.

[35] R. F. Egerton, P. Li, and M. Malac, "Radiation damage in the TEM and SEM," *Micron*, vol. 35, no. 6, pp. 399–409, Aug. 2004, doi: 10.1016/J.MICRON.2004.02.003.

[36] S. I. Wright and M. M. Nowell, "EBSD image quality mapping," *Microscopy and Microanalysis*, vol. 12, no. 1, pp. 72–84, Feb. 2006, doi: 10.1017/S1431927606060090.

[37] A. J. Schwartz, M. Kumar, B. L. Adams, and D. P. Field, "Electron backscatter diffraction in materials science," *Electron Backscatter Diffraction in Materials Science*, pp. 1–403, 2009, doi: 10.1007/978-0-387-88136-2.

[38] A. B. da S. Fanta, C. König, Y.-C. Yang, and J. R. Jinschek, "On the feasibility of using in-situ electron microscopy to investigate metal microstructures in additive manufacturing," 2024. Accessed: Sep. 29, 2025. [Online]. Available: https://orbit.dtu.dk/en/publications/on-the-feasibility-of-using-in-situ-electron-microscopy-to-invest

[39] S. Vijayan and M. Aindow, "Temperature calibration of TEM specimen heating holders by isothermal sublimation of silver nanocubes," *Ultramicroscopy*, vol. 196, pp. 142–153, Jan. 2019, doi: 10.1016/J.ULTRAMIC.2018.10.011.

[40] C. Papandrea and L. Battezzati, "A study of the α ↔ γ transformation in pure iron: rate variations revealed by means of thermal analysis," *Philosophical Magazine*, vol. 87, no. 10, pp. 1601–1618, Apr. 2007, doi: 10.1080/14786430601080260.

[41] J. Langner and J. R. Cahoon, "Increase in the alpha to gamma transformation temperature of pure iron upon very rapid heating," *Metall Mater Trans A Phys Metall Mater Sci*, vol. 41, no. 5, pp. 1276–1283, May 2010, doi: 10.1007/S11661-010-0175-9/FIGURES/13.

Supplementary material

*Influence of PFIB lift out on sample surface*

During the planar lift-out procedure, the sample surface is unavoidably exposed to high-energy Xenon ions, which may cause implantation or induce structural changes. To evaluate the influence of ion beam exposure on the microstructure, EBSD maps of an Armco iron sample (99.85 % Fe) were acquired before and after Xe-ion milling under identical conditions (2×2 binning, 8.61 ms exposure, 1 µm step size). The lift-out was performed up to the undercut stage, milling an area of 50 × 50 µm² with beam currents of 1 µA (initial milling), 0.5 µA (final milling and undercut), and 10 pA to mimic sample placement on the MEMS device. The pre-exposure EBSD scans were cropped to match the Xe-ion–exposed regions, enabling direct comparison of the microstructure before and after milling, as shown in **Figure S. 1**.

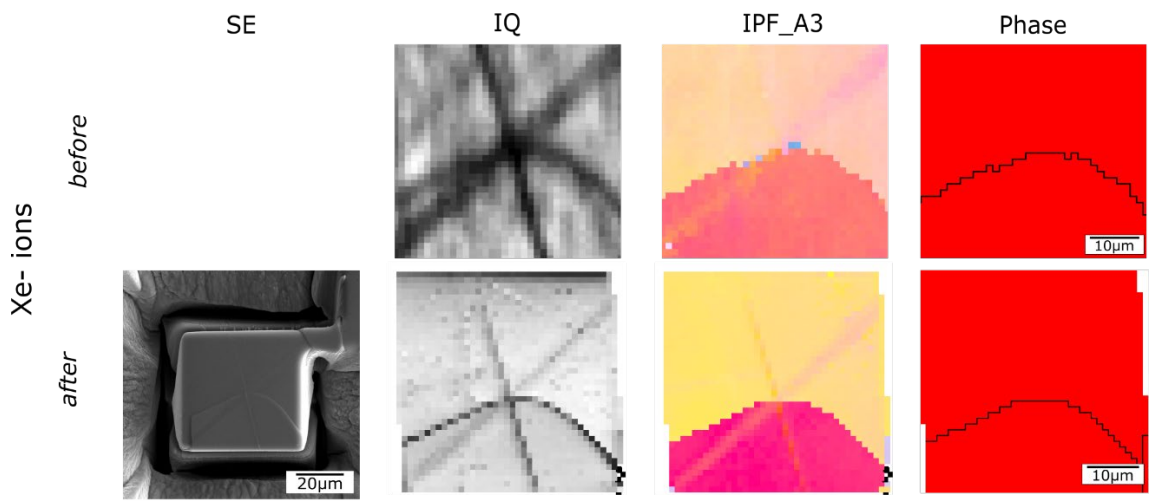

*Figure S. 1* Comparison of the effects of Xe-ion milling during the lift-out procedure on the sample surface. From left to right: secondary electron (SE) images of the milled region, EBSD image quality (IQ) maps with identical grayscale ranges before and after

*milling (cropped to the same area for comparison), and the corresponding inverse pole figure (IPF) and phase maps.*

Post-milling EBSD analysis revealed slight surface relief along grain boundaries, attributed to crystallographic channeling and increased sputtering at certain orientations. Despite this minor topography, the inverse pole figure (IPF) maps confirmed that grain boundaries and misorientations remained unchanged. A small, uniform misorientation of approximately 3° across grains was ascribed to sample realignment during transfer rather than ion-induced deformation. The improved image quality observed after Xe-ion exposure is attributed to the removal of surface oxides and shallow deformation layers rather than microstructural modification. Phase maps confirmed a fully ferritic structure, indicating that Xe-ion milling under these conditions does not induce phase transformations or crystallographic damage.

*DSC measurements*

For the Armco iron sample used in this study, the transformation temperature was verified by differential scanning calorimetry (DSC) using a Netzsch STA 449 F1 Jupiter (NETZSCH-Gerätebau GmbH, Selb, Germany). The measurement was performed at a heating rate of 10 °C min⁻¹ under an argon atmosphere.

A distinct endothermic peak was observed at 910 °C (**Figure S. 2**), confirming the ferrite–austenite transformation. The sharpness of this peak indicates that the transformation occurs rapidly and at a well-defined temperature, supporting its suitability as an internal reference for in-situ temperature calibration.

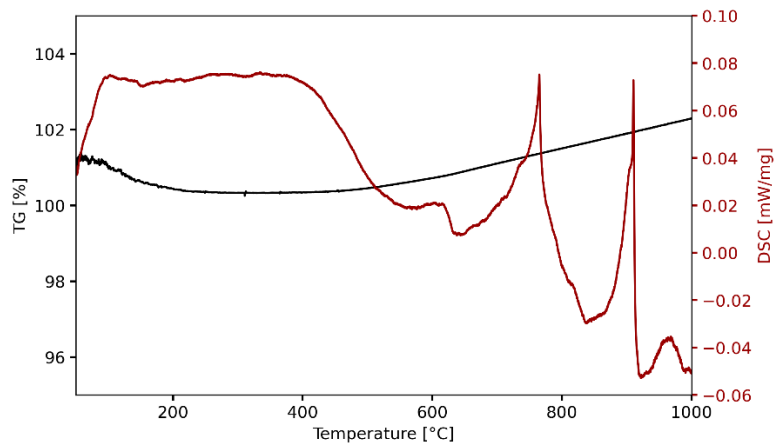

*Figure S. 2* DSC curve of the Armco iron sample during heating to 1000 °C, showing a distinct endothermic peak at 910 °C corresponding to the ferrite–austenite transformation.